\documentclass[12pt]{llncs}

\usepackage{amssymb}
\usepackage{amsmath}
\usepackage{amsfonts}

\usepackage[bookmarksopen=true,bookmarks=true,bookmarksopenlevel=1]{hyperref}
\usepackage{bookmark}
\newcommand{\doi}[1]{\textsc{doi}: \href{http://dx.doi.org/#1}{\nolinkurl{#1}}}

\newcommand{\Z}{\mathbb{Z}}
\newcommand{\R}{\mathbb{R}}

\newcommand{\Q}{\mathbb{Q}}

\newcommand{\F}{\mathbb{F}}

\usepackage{mathtools}

\DeclareMathSymbol{\sm}{\mathbin}{AMSa}{"39}

\usepackage{subcaption}
\newcommand{\overbar}[1]{\mkern 1.5mu\overline{\mkern-1.5mu#1\mkern-1.5mu}\mkern 1.5mu}
\usepackage{tikz}
\usepackage{tikz-3dplot}
\usetikzlibrary{patterns}
\usepackage{tikz-cd}
\usepackage{adjustbox}

\usepackage{tabulary}
\usepackage{algorithm}
\usepackage{algorithmicx}
\usepackage{algpseudocode}

\algnewcommand\And{\textbf{and}}
\algnewcommand\Or{\textbf{or}}
\algdef{SE}[DOWHILE]{Do}{doWhile}{\algorithmicdo}[1]{\algorithmicwhile\ #1}%

\newcommand\T{\rule{0pt}{2.6ex}}       

\usepackage{vmargin}
\setpapersize{A4}
\setmarginsrb{2.5cm}{2.0cm}{2.5cm}{2.0cm}{0pt}{1.0cm}{0pt}{1.0cm}

\tikzset{
    ncbar angle/.initial=90,
    ncbar/.style={
        to path=(\tikztostart)
        -- ($(\tikztostart)!#1!\pgfkeysvalueof{/tikz/ncbar angle}:(\tikztotarget)$)
        -- ($(\tikztotarget)!($(\tikztostart)!#1!\pgfkeysvalueof{/tikz/ncbar angle}:(\tikztotarget)$)!\pgfkeysvalueof{/tikz/ncbar angle}:(\tikztostart)$)
        -- (\tikztotarget)
    },
    ncbar/.default=0.2cm,
}
\tikzset{square left brace/.style={ncbar=0.2cm}}

\usepackage{pgfplots}
\usepackage{filecontents}
\usepackage{enumitem}

\newcommand{\algorithmfootnote}[2][\footnotesize]{%
  \let\old@algocf@finish\@algocf@finish
  \def\@algocf@finish{\old@algocf@finish
    \leavevmode\rlap{\begin{minipage}{\linewidth}
    #1#2
    \end{minipage}}%
  }%
}

\newcommand{\algcomment}[1]{%
    \vspace{3mm}%
    \noindent%
    {\footnotesize #1\par}%
    \vspace{1mm}%
    }

\title{An Implementation of the Extended Tower Number Field Sieve using 4-dimensional
Sieving in a Box and a Record Computation in $\F_{p^4}$}

\author{Ois\'in Robinson}

\institute{UCD School of Mathematics and Statistics\\ University College Dublin\\ Ireland}

\tikzcdset{every label/.append style = {font = \large}}

\usepackage{etoolbox}
\patchcmd{\abstract}{\small}{}{}{}

\begin{document}
\maketitle

\begin{abstract}
We report on an implementation of the Extended Tower Number Field Sieve (ExTNFS)
and record computation in a medium characteristic finite field $\F_{p^4}$ of 512 bits
size.  Empirically, we show that sieving in a 4-dimensional box (orthotope) for
collecting relations for ExTNFS in $\F_{p^4}$ is faster than sieving in a 4-dimensional
hypersphere. We also give a new intermediate descent method, `descent using random vectors', 
without which the descent stage in our ExTNFS computation would have been
difficult/impossible, and analyze its complexity.
\end{abstract}

\section{Introduction}\label{sec:introduction}
The number field sieve is the most efficient method known for solving the integer
factorization problem and the discrete logarithm problem in a finite field,
in the most general case.  However there are many different variants of the number field
sieve, depending on the context.  Recently, the Tower Number Field Sieve (TNFS) was
suggested as a novel approach to computing discrete logs in a finite field of
extension degree $> 1$.  The Extended Tower Number Field Sieve (ExTNFS) is a variant
of TNFS which applies when the extension degree $n$ is composite, and gives the best known
runtime complexity in the medium characteristic case (see below).\\

We briefly discuss the asymptotics of number field sieve-type algorithms.
We define the following function:
\[
L_{p^n}[\alpha,c] = \exp((c+o(1))(\log{p^n})^\alpha(\log{\log{p^n}})^{1-\alpha}).
\]
This function describes the asymptotic complexity of a subexponential function in
$\log{p^n}$, which is used to assess the complexity of the number field sieve for
computing discrete logs in $\F_{p^n}$.  For a given $p^n$, there are two important
boundaries, respectively for $\alpha = 1/3$ and $\alpha = 2/3$.  We then have 3 cases:
small characteristic, when $p < L_{p^n}[1/3,\cdot]$, medium characteristic, when
$L_{p^n}[1/3,\cdot] < p < L_{p^n}[2/3,\cdot]$, and large characteristic,
when $p > L_{p^n}[2/3,\cdot]$.  This work relates to the medium characteristic case.\\

The structure of this paper is as follows:  In Section 2, we give an overview
of computing discrete logs using ExTNFS.  In section 3, sieving in a 4-dimensional box
(orthotope) is described, and we give implementation details.  In section 4, we compare
statistics of sieving in a 4-dimensional orthotope with sieving in a 4-dimensional
hypersphere.  In section 5, we outline the
descent step, and describe our new method, including a complexity analysis.
In section 6 we give details of the record computation in $\F_{p^4}$ of 512 bits field size.
In section 7 we conclude and outline future possible work.

\section{Extended Tower Number Field Sieve (ExTNFS)}\label{sec:extnfs}

The TNFS \cite{tnfs2000},\cite{tnfs2015} makes use of a tower of number fields,
where the tower is related to a finite field extension of degree $n > 1$.
The ExTNFS \cite{MR3565295}, \cite{MR3598073} applies when $n$ is
composite.  Let $n = \eta\kappa$ where $\eta, \kappa \in \Z, \eta > 1, \kappa > 1$
(in this paper, $\eta = \kappa = 2$).  The first stage
of ExTNFS, as with all number field sieve variants, is polynomial selection.
Here, we use the `conjugation' method.  If the characteristic of our finite
field is $p$, we proceed by first choosing an irreducible polynomial $h \in \Z[y]$
of degree $\eta$,
whose reduction mod $p$, $\overbar{h}$, is irreducible in $\F_p[y]$.
Let $K = \Q(\alpha)$, where $\alpha$ is a root of $h$.
Let $\overbar{K} = \F_p(\overbar{\alpha})$, where $\overbar{\alpha}$ is a root of
$\overbar{h}$.\\

\begin{figure}
\centering
\begin{subfigure}[b]{0.4\linewidth}
\centering
\begin{tikzpicture}
	\node (Q0) at (0,-2) {$\Q$};
    \node (Q1) at (0,0) {$K=\Q[y]/\langle h\rangle$};
    \node (Q3) at (2,2) {$L_1=K[x]/\langle g_0\rangle$};
    \node (Q4) at (0,4) {$K[x]$};
    \node (Q2) at (-2,2) {$L_0=K[x]/\langle f_0\rangle$};
	\draw[<-] (Q0)--(Q1) node[midway,left] {$\eta$};
    \draw[<-] (Q1)--(Q2) node[midway,below left] {$\kappa$};
    \draw[<-] (Q3)--(Q4);
    \draw[<-] (Q2)--(Q4);
    \draw[<-] (Q1)--(Q3) node[midway,below right] {$2\kappa$};
\end{tikzpicture}
\caption{Commutative diagram of ExTNFS.}
\end{subfigure}
\hspace{20mm}
\begin{subfigure}[b]{0.4\linewidth}
\centering
\begin{tikzpicture}
	\node (Q0) at (0,-1.7) {$\F_p$};
    \node (Q1) at (0,1) {$\overbar{K}=\F_p[y]/\langle \overbar{h}\rangle$};
    \node (Q2) at (0,3.7) {$\overbar{L_0}=\overbar{K}[x]/\langle \overbar{f_0}\rangle$};
	\draw[<-] (Q0)--(Q1) node[midway,left] {$\eta$};
    \draw[<-] (Q1)--(Q2) node[midway,left] {$\kappa$};
\end{tikzpicture}
\caption{Finite Field Tower}
\end{subfigure}
\label{fig:tnfscommute}
\end{figure}

Let $\mathfrak{p}$ be a prime ideal in the ring of integers of $K$ lying above $p$.
Let $f_0 \in K[x]$ be irreducible of degree $\kappa$ such that $\mathfrak{p}$ does not divide
the leading coefficient of $f_0$,
and such that its reduction mod $\mathfrak{p}$, $\overbar{f_0}$, is irreducible in $\overbar{K}[x]$.
Let $\beta$ be a root of $f_0$ and let $L_0=K(\beta)$.
Let $g_0 \in K[x]$ be irreducible of degree $2\kappa$  such that its reduction mod $\mathfrak{p}$, $\overbar{g_0}$, 
shares an irreducible factor of degree $\kappa$ with $\overbar{f_0}$.
Let $\gamma$ be a root of $g_0$ and let $L_1=K(\gamma)$.
Let $F = \F_{p^{\eta\kappa}} = (\F_p(\overbar{\alpha})[x])/\langle \overbar{f_0}\rangle = \overbar{K}[x]/\langle \overbar{f_0}\rangle$.\\

To find $f_0$ and $g_0$ in practice we recall the following algorithm.
Assume $s \in \Z$ is a square mod $p$ but not a square in $\Z$. Let $r$ be a square root of $s$ mod $p$.
Then choose $t \in \Z[\alpha]$, with its reduction $\overbar{t} \in \F_p(\overbar{\alpha})$.
Typically, $t$ is constant or a linear polynomial in $\alpha$. Let
\begin{align*}
g_0	&= t^2x^{2\kappa} + \left(2t^2 - s\right)x^\kappa + t^2\,
\end{align*}
ensuring $g_0$ is irreducible over $K$ and then
\begin{align*}
\overbar{g_0} &= \left( \overbar{t}x^\kappa+rx^{\kappa/2}+\overbar{t} \right)\left( \overbar{t}x^\kappa-rx^{\kappa/2}+\overbar{t} \right)\\
\end{align*}
Assume $\overbar{g_0}$ has an irreducible factor of degree $\kappa$.\\

With $s,r$ chosen as above, let $L$ be the lattice with basis
\begin{align*}
\begin{bmatrix}
p & & r \\
0 & & 1 
\end{bmatrix}
\end{align*}
and find a short vector $(u,v)^T$ in $L$.  Finally, let
\[
f_0 = tvx^\kappa+ux^{\kappa/2}+tv,
\]
making sure $f_0$ is irreducible over $K$.  Define $g_0$ as above.  Then
\begin{align*}
L_0 &\cong K[x]/\langle f_0\rangle \\
L_1 &\cong K[x]/\langle g_0\rangle.
\end{align*}
Note that $L_0$ and $L_1$ then share a built-in automorphism of order 2.
This completes the selection of $f_0$, $g_0$ by the conjugation method.
However, the quality of polynomial pairs varies.  We can either examine
the alpha values as per \cite{guillevic:hal-02263098}, or conduct sieving tests, to
find the best pair.\\

After polynomial selection comes the sieving stage.  In this stage
we look for relations between $L_0$ and $L_1$ coming from the same element
of $K[x]$ and which simultaneously have absolute norm divisible only
by primes up to some fixed bound $B$.  In fact we construct a factor
base of prime ideals consisting of all prime ideals, of any possible degree,
in both $L_0$ and $L_1$, of norm up to $B$, and if we require (rational)
primes dividing the norm to be less than or equal to $B$, we can be certain
our factor base is sufficient to cover any possible ideal factorization.
In practice we focus mostly on degree 1 ideals, allowing a small number
of degree 2 ideals so that we can keep as many relations as possible.\\

The optimum sieving dimension $d$ depends on the target field, but for the Extended
Tower variants the dimension must be $\ge 4$.  We find the optimum $d$ by
a consideration of the average norms from typical sieving ideals.
Once $d$ is fixed we have another consideration as to how to conduct the
sieving in practice.  It was shown in \cite{GDeMthesis} that asymptotically
sieving in a $d$-sphere produces the smallest norms on average, but it is
an interesting question to see how large $d$ can be and still have fast
sieving in a $d$-dimensional orthotope, as in certain low dimensions we might
be able to exploit the shape of the sieving planes to allow faster enumeration.
We describe an approach to 4d sieving in the next section which was used
in the record computation detailed in section \ref{Record}.  Our timings
indicate that 4d sieving in an orthotope is practical and fast.\\

The question of how many relations are sufficient to conclude the sieving
stage is quite delicate, and this can make or break the entire computation.
This phenomenon is complicated
by the fact that a key post-processing step after the sieving stage is the
duplicate removal step.  In the ExTNFS setting, there are a number
of different types of duplicates which must be removed to prevent problems
in the subsequent stages.  However, removing all types of duplicates tends
to drastically reduce the number of relations, possibly even below the number
of ideals (unknowns), thereby making the linear algebra stage intractable.
We discuss this in more detail in section \ref{Record}, but for now
we note that we must aim to have more relations than ideals in our relation set,
even after duplicate removal.\\

The duplicate removal step for ExTNFS was described comprehensively in
\cite{GDeMthesis}.  We find experimentally that if the different types
of duplicates are not removed, the matrix preparation routines such
as purge (the removal of singletons) and merge (Gaussian elimination
to reduce the size of the matrix) tend to fail/not function correctly.\\

The relation set is then reduced using singleton removal and a limited
amount of Gaussian elimination.  At this stage for each row of the matrix,
depending on the unit rank of $L_0$ and $L_1$ we augment the matrix
with Schirokauer maps, which are linear maps that deal with the
problem of units when attempting to obtain a vector in the right nullspace
of the matrix corresponding to the virtual logs of ideals.
The usual algorithm for finding the nullspace vector is the Block
Wiedemann algorithm, as it allows distribution of the computation
across the nodes of a cluster.\\

The solution vector from the linear algebra step is then used to
create a database of ideal virtual logs in the log reconstruction step.
Finally for a given generator and target element of $F$ we must
carry out the descent step, where we relate the virtual logs of
generator and target (lifted to $L_0$ or $L_1$) to known virtual
logs in the database.  There are various methods for doing this.
We describe the descent in more detail in section \ref{Descent}.\\

\section{4d Sieving in a Box}\label{section3}
Using the previous notation, suppose our intermediate number field is $K=\Q(\alpha) \cong \Q[y]/\langle \overbar{h} \rangle$,
and our sieving number field $L_0$ is defined by the polynomial $f_0 \in K[x]$.
We can compute the defining polynomial $f$ of the absolute field of $L_0$ (i.e. not a relative
extension).  We have
\[
f = \text{Res}_\alpha{\left(f_0, h\right)}
\]
where $\text{Res}_\alpha$ means the polynomial resultant with respect to $\alpha$.\\

Suppose we have a special-q rational prime $q$.  We want to find principal ideals of the
ring of integers of $L_0$ that are divisible by a prime $\mathfrak{q}$ above $q$
in $L_0$, and by many smaller prime ideals.  The most common case is the degree-1 case.
Find $r$ and $R$ such that
\[
h(r) \equiv 0\,\mod{q}
\]
and
\[
f(R) \equiv 0\, \mod{q}.
\]
We can think of our degree-1 sieving ideal $\mathfrak{q}$ above $q$ as being represented
uniquely by the triple $(q,r,R)$.\\

In general there are a number of possible ideal degrees (inertia degrees) for 
$\mathfrak{q}$.  The inertia degree divides the degree of the absolute defining
polynomial $f$, however not necessarily all divisors are present.  In our record
computation, the polynomial $f_0$ on side 0 has degree 4 and the polynomial $g_0$
on side 1 has degree 8, but e.g. on side 1 there only exist ideals of degree 1,2,4.
This is an artefact of the conjugation method of polynomial selection.\\

We outline the method when the conjugation method has produced $h$ of degree 2,
$f_0$ with $f = \text{Res}_\alpha\left(f_0,h\right)$ of degree 4 and 
$g_0$ with $g = \text{Res}_\alpha\left(g_0,h\right)$ of degree 8, for ExTNFS in the
$\F_{p^4}$ setting.\\

There are 3 cases for the inertia degree for our 4d sieve, for each of which we 
form a lattice basis.  We split the degree 2 case into two further cases.

\subsection{Degree 1}
Let
\begin{equation*}
\begin{split}L=
\begin{bmatrix}
\phantom{-}q & -r & -R & \phantom{-}0\phantom{-}\\
\phantom{-}0 & \phantom{-} 1 & \phantom{-}0 & -R\phantom{-}\\
\phantom{-}0 & \phantom{-} 0 & \phantom{-}1 & \phantom{-}0\phantom{-}\\
\phantom{-}0 & \phantom{-} 0 & \phantom{-}0 & \phantom{-}1\phantom{-}
\end{bmatrix},\hspace{7mm}
\end{split}
\begin{split}
h(r) \equiv 0 \mod{q}\\
f(R) \equiv 0 \mod{q}
\end{split}
\end{equation*}

\subsection{Degree 2 type 1}
Let
\begin{equation*}
\begin{split}L=
\begin{bmatrix}
\phantom{-}q & -r & \phantom{-}0 & \phantom{-}0\phantom{-}\\
\phantom{-}0 & \phantom{-}1 & \phantom{-}0 & \phantom{-}0\phantom{-}\\
\phantom{-}0 & \phantom{-}0 & \phantom{-}q & -r\phantom{-}\\
\phantom{-}0 & \phantom{-}0 & \phantom{-}0 & \phantom{-}1\phantom{-}
\end{bmatrix},\hspace{7mm}
\end{split}
\begin{split}
h(r) \equiv 0 \mod{q}
\end{split}
\end{equation*}
and $f$ has no roots mod $q$.

\subsection{Degree 2 type 2}
Let
\begin{equation*}L=
\begin{bmatrix}
\phantom{-}q & \phantom{-}0 & \phantom{-}a_0 & \phantom{-}0\phantom{-} \\
\phantom{-}0 & \phantom{-}q & \phantom{-}a_1 & \phantom{-}a_0\phantom{-} \\
\phantom{-}0 & \phantom{-}0 & \phantom{-}1  & \phantom{-}a_1\phantom{-} \\
\phantom{-}0 & \phantom{-}0 & \phantom{-}0  & \phantom{-}1\phantom{-}
\end{bmatrix}
\end{equation*}
where $h$ is irreducible mod $q$, $f$ has no roots mod $q$ and
\begin{equation*}
\begin{split}
& (x + a_1\alpha + a_0)\vert f_0 \mod{\mathfrak{q}}.
\end{split}
\end{equation*}
Note that this requires an implementation of bivariate polynomial factorization.

\subsection{Degree 4}
Let
\begin{equation*}L=
\begin{bmatrix}
\phantom{-}q & \phantom{-}0 & \phantom{-}0 & \phantom{-}0\phantom{-} \\
\phantom{-}0 & \phantom{-}q & \phantom{-}0 & \phantom{-}0\phantom{-} \\
\phantom{-}0 & \phantom{-}0 & \phantom{-}q & \phantom{-}0\phantom{-} \\
\phantom{-}0 & \phantom{-}0 & \phantom{-}0 & \phantom{-}q\phantom{-}
\end{bmatrix}
\end{equation*}
where $q$ is inert in $L_0$.\\

In practice we only sieve degree 1 ideals, unless the reduced basis has very small
coefficients.  Note that when $q$ is not very small, the determinant of the degree 2
basis is already $q^2$ and is not likely to have vectors of small norm.\\

We reduce $L$ using e.g. the LLL algorithm.  This is the basis of our sieving lattice.
We must enumerate all points in the lattice contained in a 4d orthotope, of shape
$\mathcal{S} =[-B,B[\times[-B,B[\times[-B,B[\times[-B,B[$, for some bound $B$.  The enumeration
is simple in principle:  We enumerate in lines, then planes, then hyperspaces.

Suppose the sieving lattice has basis
\begin{equation*}
\begin{split}L=
\begin{bmatrix}
\phantom{-}u_1 & \phantom{-}v_1 & \phantom{-}w_1 & \phantom{-}t_1 \phantom{-}\\
\phantom{-}u_2 & \phantom{-}v_2 & \phantom{-}w_2 & \phantom{-}t_2 \phantom{-}\\
\phantom{-}u_3 & \phantom{-}v_3 & \phantom{-}w_3 & \phantom{-}t_3 \phantom{-}\\
\phantom{-}u_4 & \phantom{-}v_4 & \phantom{-}w_4 & \phantom{-}t_4 \phantom{-}
\end{bmatrix}.
\end{split}
\end{equation*}
We compute the 4d normal $N = (n_x,n_y,n_z,n_t)$ to the 3d hyperspace determined by the first
3 (column) vectors of $L$, using the cross product, as
\begin{align*}
n_x &= (-w_4v_3 + w_3v_4)u_2 + (\phantom{-}w_4v_2 - w_2v_4)u_3 + (-w_3v_2 + w_2v_3)u_4\\
n_y &= (\phantom{-}w_4v_3 - w_3v_4)u_1 + (-w_4v_1 + w_1v_4)u_3 + (\phantom{-}w_3v_1 - w_1v_3)u_4\\
n_z &= (-w_4v_2 + w_2v_4)u_1 + (\phantom{-}w_4v_1 - w_1v_4)u_2 + (-w_2v_1 + w_1v_2)u_4\\
n_t &= (\phantom{-}w_3v_2 - w_2v_3)u_1 + (-w_3v_1 + w_1v_3)u_2 + (\phantom{-}w_2v_1 - w_1v_2)u_3
\end{align*}
We start at the point $V_T = (0,0,0,0)$ which is always contained in
$\mathcal{S}$.
We keep subtracting the vector $T = (t_1,t_2,t_3,t_4)$ from $V_T$ until
we cannot subtract $T$ any more and still have $\mathcal{S_N} \cap \mathcal{S} \neq \emptyset$,
where $\mathcal{S_N}$ is the 3d subspace determined by $N$ and $V_T$.
Thus we relocate $V_T$ until we are at the start of a sequence of 3d subspaces obtained
by adding $T$ to $V_T$.  We then enumerate within this 3d subspace.  When this is
finished, we add $T$ to $V_T$ and start enumerating from $V_T$ within the second
3d subspace, and so on until the 3d subspace no longer intersects $\mathcal{S}$.\\

The rest of the enumeration proceeds in an exactly analogous way, except in lower
dimensions.  We require a subroutine for determining if a subspace intersects
$\mathcal{S}$.  Algorithm \ref{alg1} shows a method to determine
if the 3d subspace $\mathcal{S_N}$ with 4d normal $(n_x,n_y,n_z,n_t)$ containing the point
$V = (x,y,z,t)$ intersects the 4d orthotope $\mathcal{S}$.
This algorithm works by using the dot product between a vector and the normal $N$ to
determine which `side' of $\mathcal{S_N}$ a point is on.  If $V$ is always
on the same side as the respective vertices of the boundary, it follows that 
$\mathcal{S_N} \cap \mathcal{S} = \emptyset$, otherwise the converse is true.\\

The `enumeration in lower dimension' has been detailed in \cite{MR4178778} for
dimension 3.  We just need to add an extra co-ordinate to any vectors used, but otherwise
the algorithm is practically identical.  For example we show a 4d version of the integer
linear programming algorithm (Algorithm 2, \cite{MR4178778}) in Algorithm \ref{alg2}.\\

An implementation note:  The algorithm speed still seems to come mostly from fast enumeration
in a plane.  Moving from plane to plane is done less frequently, and switching
3d subspace is done even less frequently, so any code complexity is offset by this.
Note also that in our implementation, we allow orthotopes of `rectangular' shape
by allowing nonequal sides, we gave a single side length of $B$ above for the sake
of simplicity.

\begin{algorithm}[!ht]
\caption{3d Subspace Intersects Box} 
\begin{algorithmic}
	\Require
		\State \hspace{5mm}Boundary $\mathcal{S} = [-B,B[\times[-B,B[\times[-B,B[\times[-B,B[$
		\State \hspace{5mm}4d normal $N=(n_x,n_y,n_z,n_t)$ to 3d subspace $\mathcal{S_N}$
		\State \hspace{5mm}Point $V=(x,y,z,t)$ contained in $\mathcal{S_N}$
	\Ensure
		True if $\mathcal{S_N} \cap \mathcal{S} \neq \emptyset$, False otherwise
	\vspace{1mm}
\Procedure{}{}
    \vspace{1mm}
    \State $d \gets n_xx + n_yy + n_zz + n_tt$
    \State $nxB0 \gets -n_xB$
    \State $nyB0 \gets -n_yB$
    \State $nzB0 \gets -n_zB$
    \State $ntB0 \gets -n_tB$
    \State $nxB1 \gets n_x(B-1)$
    \State $nyB1 \gets n_y(B-1)$
    \State $nzB1 \gets n_z(B-1)$
    \State $ntB1 \gets n_t(B-1)$
    \State $s_0 \gets nxB0 + nyB0 + nzB0 + ntB0 - d$
    \State $s_1 \gets nxB1 + nyB0 + nzB0 + ntB0 - d$
    \State $s_2 \gets nxB0 + nyB1 + nzB0 + ntB0 - d$
    \State $s_3 \gets nxB1 + nyB1 + nzB0 + ntB0 - d$
    \State $s_4 \gets nxB0 + nyB0 + nzB1 + ntB0 - d$
    \State $s_5 \gets nxB1 + nyB0 + nzB1 + ntB0 - d$
    \State $s_6 \gets nxB0 + nyB1 + nzB1 + ntB0 - d$
    \State $s_7 \gets nxB1 + nyB1 + nzB1 + ntB0 - d$
    \State $s_8 \gets nxB0 + nyB0 + nzB0 + ntB1 - d$
    \State $s_9 \gets nxB1 + nyB0 + nzB0 + ntB1 - d$
    \State $s_a \gets nxB0 + nyB1 + nzB0 + ntB1 - d$
    \State $s_b \gets nxB1 + nyB1 + nzB0 + ntB1 - d$
    \State $s_c \gets nxB0 + nyB0 + nzB1 + ntB1 - d$
    \State $s_d \gets nxB1 + nyB0 + nzB1 + ntB1 - d$
    \State $s_e \gets nxB0 + nyB1 + nzB1 + ntB1 - d$
    \State $s_f \gets nxB1 + nyB1 + nzB1 + ntB1 - d$
	\State If all of $s_0,\dots,s_f$ have the same sign, return False, otherwise return True
\EndProcedure
\end{algorithmic}\label{alg1}
\end{algorithm}

\begin{algorithm}[!ht]
\caption{Integer Linear Programming for 3d Subroutine of 4d Sieve} 
\begin{algorithmic}
	\Require
		\State \hspace{5mm}Boundary $[-B,B[\times[-B,B[\times[-B,B[\times[-B,B[$,
		\State Plane $P$ defined by $u=(u_1,u_2,u_3,u_4),v=(v_1,v_2,v_3,v_4),R=(x,y,z,t)$
		\State \hspace{5mm} such that $R \in P$, but $R$ not necessarily in boundary.
	\Ensure
		\State \hspace{5mm}$(a,b)$ such that $p_0 = R + a\cdot u + b\cdot v$ contained in
		both $P$ and boundary.
\Procedure{}{}
    \vspace{1mm}
	\State $U \gets \{ u_1,u_2,u_3,u_4,-u_1,-u_2,-u_3,-u_4 \}^T$
	\State $V \gets \{ v_1,v_2,v_3,v_4,-v_1,-v_2,-v_3,-v_3 \}^T$
	\State $C \gets \{ B-x-1,B-y-1,B-z-1,B-t-1,B+x,B+y,B+z,B_t \}^T$
	\State $L \gets \lvert \text{LCM}\left(u_1,u_2,u_3,u_4\right) \rvert$
	\State Normalize system - multiply all entries by $L$, divide $U_i,V_i,C_i$ by original $\lvert U_i \rvert$
	\State $a \gets B, b \gets B$
	\For {all pairs of rows $(i,j)$}
		\If {$U_i = 0$ and $V_i > 0$}
			\State $b_{_{TRIAL}} = \left\lfloor C_i / V_i \right\rfloor$
			\State if $b_{_{TRIAL}} < b$ then $b \gets b_{_{TRIAL}}$
		\Else
			\If {$U_i < 0$ and $U_j > 0$ and $\lvert i - j \rvert \neq 4$ }
				\State $D \gets V_i + V_j$
				\If {$D > 0$}
					\State $b_{_{TRIAL}} = \left\lfloor \left(C_i + C_j\right)/D \right\rfloor$
					\State if $b_{_{TRIAL}} < b$ then $b \gets b_{_{TRIAL}}$
				\EndIf
			\EndIf
		\EndIf
	\EndFor
	\State for the best value of $b$, compute $a$ by back substitution.
	\State Return $(a,b)$
    \State\EndProcedure
\end{algorithmic}\label{alg2}
\end{algorithm}

\section{Comparison of sieving in sphere versus box}\label{chap8}

In this section we report on some statistics collected on various timings of
our implementations of sphere sieving and orthotope sieving.  Our 3d orthotope
sieve was originally used in a record computation (at the time) of computing
discrete log in a finite field of shape $\F_{p^6}$ and size 423 bits, using
the traditional number field sieve.  The $\F_{p^6}$ computation has since been
superseded by a record computation of G.DeMicheli, P.Gaudry, C.Pierrot in a
finite field $\F_{p^6}$ of size 521 bits using exTNFS \cite{GDeMpaper}.
Our 4d orthotope sieve was subsequently used in a record discrete log computation
in a finite field $\F_{p^4}$ of size 512 bits, using exTNFS.  We give details of
these computations in section \ref{Record}.  The purpose of this section is to focus
solely on  the sieving aspect in 4 dimensions, and compare the box versus sphere
approach. We implemented G.DeMicheli's sphere sieve in C\texttt{++} in 4d
based on the description in \cite{GDeMthesis}.

\subsection{Evaluation methodology}\label{evalmethod}

The ultimate test of a sieving method is how many (unique) relations per second it is
capable of producing.  In this sense, all other considerations are moot.
This single metric determines how fast the sieving stage of NFS or TNFS runs, which
is the bottleneck of the algorithm.  However it is also academically interesting
to compare other statistics of the box versus sphere approach to sieving, such as
number of points enumerated in a given time, percentage of all points hit etc. and 
their variations with e.g. field shape, special-q size, sieving side etc.\\

\noindent G.De Micheli showed \cite{GDeMthesis} that asymptotically,
sieving in an $n$-dimensional hypersphere, i.e. enumerating all
vectors in such, produces vectors that on average have smaller norm
than vectors from exhaustive enumeration in an $n$-dimensional
orthotope.  Asymptotically, this is a great advantage since
a vector is more likely to produce a relation if the norm (which is an
integer) is smaller in absolute value.  It is not initially clear in which dimension
we lose any possible advantage gained from sieving in a box (if any).
The only real way to know for sure is to implement both and compare
runtime and other statistics, so this is what we do.\\

\noindent In what follows, we implemented each sieve in dimension 4 in C\texttt{++},
compiling with the Gnu compiler version 9.4.0.  The programs were run
on the Kay cluster at the Irish Centre for High-End Computing (ICHEC),
where each node has 2 $\times$ 20-core 2.4 GHz
Intel Xeon Gold 6148 (Skylake) processors and 192GB of RAM.

\subsection{4d sieve in box versus sphere}
For the case of 4d sieving, with 4 being the first composite integer, we used our 4d box sieve
written for exTNFS to solve the discrete log in $\F_{p^4}$ where
$p^4$ has 512 bits (the current record in degree 4 extensions).
We gathered runtime statistics for the same setup as our record computation and present our
findings below.\\

\noindent We used two different measures in comparing 4d sieving in a box versus sphere.
The first is `yield per million points enumerated' (yield ppm).  This is the number
of relations produced per million points enumerated per core.  This places more emphasis
on the efficiency of the sieve enumeration at producing unique relations.  The second
measure is simply the number of unique relations produced per core hour.  This takes into
account all timings in the production of relations, including enumeration time,
histogram time, cofactorization time and any other lesser time costs.\\

\noindent Our finding is that in 4d, sieving in a box significantly outperforms
sieving in a sphere in three ways:  First:
\begin{remark}
With optimal geometry, the box sieve
has significantly higher relations per core hour (see tables \ref{tab4dbox},
\ref{tab4dsphere} and figure \ref{figrch}).
\end{remark}

\noindent Second:
\begin{remark}
In all competitive configurations of geometry it is possible to have
a smaller histogram size in the box case, where for sphere sieving it usually becomes
necessary to use a 4GB histogram (as a result of coordinates being in
$[-R,R]^4$ where $R$ is at least 65).  In practice, this considerably limits
the number of cores that can sieve simultaneously on a production compute node.
The nodes we used had 40 cores and 192GB of RAM, and to produce tables \ref{tab4dbox}
and \ref{tab4dsphere}
we restricted both box and sphere cases to 30 cores out of 40, otherwise the sphere
sieve would have run out of memory.  There was no such limitation in the box
case, so in fact at 75\% node capacity, the box sieve still outperforms the sphere
sieve.
\end{remark}

\noindent Third:
\begin{remark}\label{yieldppm}
Enumeration speed is slightly higher in the box case in 4d.
We also use the measure `yield parts per million' to describe the number of
relations produce per million points enumerated, but it is important to note
that this is independent of the time taken.
While sphere sieving has higher yield ppm than box sieving in 4d, and
technically it is producing a greater fraction of valid unique relations
in its sieving region, it does not do this fast enough.
\end{remark}

\begin{remark}
For our 512-bit DLP computation in $\F_{p^4}$ the box sieve had even
another advantage, which was some additional sieving of degree 2 ideals which 
increases the relations per core hour slightly, at very little extra cost.
\end{remark}

\begin{table}
\begin{center}
\begin{tabulary}{13cm}{|C|C|C|C|}
\hline
\T Label&geometry of sieve region&histogram size(bytes)&rels/core hour\\
\hline
\T A&$2^8\times2^8\times2^7\times2^7$&1073741824&531\\
B&$2^8\times2^7\times2^7\times2^7$&536870912&1057\\
C&$2^7\times2^7\times2^7\times2^7$&268435456&1019\\
D&$2^7\times2^7\times2^7\times2^6$&134217728&834\\
E&$2^7\times2^7\times2^6\times2^6$&67108864&674\\
F&$2^7\times2^6\times2^6\times2^6$&33554432&485\\
G&$2^6\times2^6\times2^6\times2^6$&16777216&352\\
\hline
\end{tabulary}
\vspace{1mm}
\caption{exTNFS, 4d sieving $\F_{p^4}$, box, 80 cores for 1 hour}\label{tab4dbox}
\end{center}
\end{table}

\begin{table}
\begin{center}
\begin{tabulary}{13cm}{|C|C|C|C|}
\hline
\T Radius of sieve sphere&coordinate bits&histogram size(bytes)&rels/core hour\\
\hline
\T 105&8&4294967296&760\\
100&8&4294967296&802\\
95&8&4294967296&786\\
90&8&4294967296&755\\
85&8&4294967296&714\\
80&8&4294967296&682\\
75&8&4294967296&636\\
70&8&4294967296&571\\
65&8&4294967296&497\\
60&7&2147483648&637\\
\hline
\end{tabulary}
\vspace{1mm}
\caption{exTNFS, 4d sieving $\F_{p^4}$, sphere, 80 cores for 1 hour}\label{tab4dsphere}
\end{center}
\end{table}

\pgfplotsset{scaled y ticks=false}
\pgfplotsset{scaled x ticks=false}
\begin{figure}
\begin{center}
\begin{tikzpicture}
  \begin{axis}[
  	/pgf/number format/set thousands separator = {},
    xlabel = 4d Sieve geometry label,
    ylabel = Relations/core hour,
	ylabel style={yshift = 5mm},
	ytick distance = 100,
	xtick = data,
	xticklabels = {A,B,C,D,E,F,G}
    ]
    \addplot [only marks, blue] table[x index=0,y index=1,header=false] {data1.csv};
    \addplot [only marks, red] table[x index=0,y index=1,header=false] {data2.csv};
	\label{p2}
    \addplot [no markers, gray] gnuplot [raw gnuplot] { 
            f(x) = 1065*1.75*(x-0.51)*2^(-0.75*(x-0.11)); 
			a=1.75; b=0.75;
            plot [x=1:7] f(x); 
    };
    \addplot [no markers, gray] gnuplot [raw gnuplot] { 
            g(x) = 385+800*0.87*(x+0.1)*2^(-0.7*(x+0.6)); 
			a=1.75; b=0.75;
            plot [x=1:7] g(x); 
    };
    \legend{box}
  \end{axis}
  \node [draw,fill=white] at (rel axis cs: 0.95,0.87) {\shortstack[l]{
  \ref{p2} sphere}};
\end{tikzpicture}
\end{center}
\caption{exTNFS, relations/core hour, box vs sphere 4d sieve}\label{figrch}
\end{figure}

\begin{remark}
Each point enumerated involves a trip to main memory to store a point (encoded as a
32-bit integer), since the number of points is large and quickly goes beyond anything
that might benefit from the various levels of cache.  Therefore the number of points
enumerated is closely related to the total enumeration time.
\end{remark}

\begin{remark}
We would like to make a fair comparison between box and sphere-based sieving.  One
way of comparing is to arrange for the box and sphere to have the same volume, 
since the number of lattice points contained in a volume $V$ depends closely on $V$.
For instance, say the box in 4d has side length $2^7 = 128$, so the volume
is $2^{28} = 268435456$.  We can compute the radius $R$ of the hypersphere that 
has the same volume using the formula for the volume of a 4d hypersphere.
This gives $R = 85.88$ approximately, and so a 
4d hypersphere with this radius should contain very nearly the same number
of lattice points as the box.  However note that as the dimension increases,
the ratio of the volume of the hypersphere to that of the box decreases significantly,
in fact as the dimension tends to infinity, the volume of the hypersphere tends to zero.
Since it is not obvious in which dimension the advantage of asymptotically lower norms
from sieving in a hypersphere first becomes apparent in these low dimensions,
all we can do is conduct tests to find the `sweet spot' of enumeration speed/yield ppm.
\end{remark}

\begin{remark}
Notice in tables \ref{tab4dcomp1} and \ref{tab4dcomp2} with regard to the yield ppm for
the box and sphere case, that even though more points were enumerated for a given
special-q range in the box case, the sphere sieve produces significantly more relations,
a surplus of about 25\% more.  The explanation is that the effect of smaller average
norm in the sphere geometry is already apparent in 4d, and more relations are found,
again (in the 4d case) just not fast enough.
\end{remark}

\begin{remark}
The selection of optimal parameters is quite subtle, for example naively choosing
the largest box or sphere for which our vectors are encoded in at most 32 bits
may be suboptimal.  We may (sometimes do) find that a smaller box gives a higher
unique relation production rate, which is the only metric that matters in the sieving
stage.
\end{remark}

\begin{remark}
The relatively low memory cost of the box sieve indicates the possibility
for a particularly fast GPU-based implementation of the 4d lattice sieve, since
usually the impediment to a GPU sieve for the number field sieve is the high
memory cost.
\end{remark}

\begin{table}
\begin{center}
\begin{tabulary}{13cm}{|l|C|C|}
\hline
\T Shape of 4d region&box&sphere\\
\hline
\T Geometry of sieve region&$2^7\times2^7\times2^7\times2^6$&radius $80$\\
\#special-q&1000&1000\\
Min special-q&10000189&10000189\\
Max special-q&12001189&12001189\\
Special-q side&0&0\\
\hline
\T Sieving side&0&0\\
Av. \#points in sieve region per q&129745888&101769710\\
Av. \#points enumerated per q&108137268&101769710\\
Percentage of points hit&83.35&100\\
Av. enumeration time (s)&7.83&7.51\\
Av. points enumerated per ms per q&13819&13556\\
\hline
\T Sieving side&1&1\\
Av. \#points in sieve region per q&127391309&99922910\\
Av. \#points enumerated per q&106189065&99922910\\
Percentage of points hit&83.36&100\\
Av. enumeration time (s)&7.77&7.45\\
Av. points enumerated per ms per q&13662&13409\\
\hline
\T Relation count&6965&8707\\
Unique relation count&5729&7585\\
Yield ppm (but see remark \ref{yieldppm})&53.95&75.91\\
\hline
\end{tabulary}
\end{center}
\vspace{1mm}
\caption{exTNFS, 4d sieving $\F_{p^4}$, box vs sphere, q=10000189-12001189}\label{tab4dcomp1}
\end{table}

\begin{table}
\begin{center}
\begin{tabulary}{13cm}{|l|C|C|}
\hline
\T Shape of 4d region&box&sphere\\
\hline
\T Geometry of sieve region&$2^7\times2^7\times2^7\times2^6$&radius $80$\\
\#special-q&1000&1000\\
Min special-q&37503223&37503223\\
Max special-q&39503953&39503953\\
Special-q side&0&0\\
\hline
\T Sieving side&0&0\\
Av. \#points in sieve region per q&129721890&101773958\\
Av. \#points enumerated per q&107483024&101773958\\
Percentage of points hit&82.86&100\\
Av. enumeration time (s)&7.97&7.67\\
Av. points enumerated per ms per q&13484&13266\\
\hline
\T Sieving side&1&1\\
Av. \#points in sieve region per q&127381471&99925799\\
Av. \#points enumerated per q&104866866&99925799\\
Percentage of points hit&82.33&100\\
Av. enumeration time (s)&7.91&7.61\\
Av. points enumerated per ms per q&13253&13124\\
\hline
\T Relation count&4790&6104\\
Unique relation count&4064&5422\\
Yield ppm&45.68&54.26\\
\hline
\end{tabulary}
\end{center}
\vspace{1mm}
\caption{exTNFS, 4d sieving $\F_{p^4}$, box vs sphere, q=37503223-39503953}\label{tab4dcomp2}
\end{table}

\subsection{Summary}
Experimentally we find that sieving in a box in 4d produces significantly more relations
per core hour than sieving in a sphere, with respectively optimal configuration.
In our tests, sieving in a box in 4d for a field size of 512 bits produced over
1000 relations/core hour, whereas sieving in a sphere produced only just over 800
relations/core hour.  However we also had to restrict the sphere sieve to 30 cores
out of 40 on a node, due to the memory constraint - there is no such restriction
necessary in the 4d box case.

\scalebox{0.85}{
\begin{minipage}{1.2\linewidth}
\begin{algorithm}[H]
\caption{Special-$\mathfrak{q}$ Descent for ExTNFS$^\dagger$} 
\begin{algorithmic}[1]
	\Require
		\Statex Let $L_0, L_1$ be the number field towers of ExTNFS.
		\Statex Given generator, target $g, t \in F = \F_{p^n}$, group order $\ell$.
		\Statex Given a database of logs of ideals with prime norm up to $B$.
		\Statex Assume the virtual log of $g$ is known.
	\Ensure
		\Statex Produce the log of $t$ to the base $g$ in $F$.
    \vspace{1mm}
	\State Choose a bound $B_L > B$ for large primes.
	\State Using the algorithm of Guillevic \cite{GuillevicDescent}, find $i \in \Z$ such that
	\State $h$, the lift of $tg^i$ to (say) $L_0$, has norm which factors into primes bounded by $B_L$.
	\State Choose an intermediate bound $B_I$ with $B < B_I < B_L$.
	\For {Each prime $q > B_I$ in the factorization of $\mathcal{N}(h)$}
		\State Construct the sieving lattice $L$ as shown in section 3
		\State Reduce $L$ with LLL, giving $\overbar{L}$.
		\Do 
			\State Produce a random vector in $\overbar{L}$
			\State Let $\overbar{h}$ be the corresponding element of $F$
			\State Let $h_0,h_1$ be the lifts of $\overbar{h}$ to $L_0,L_1$ resp.
		\doWhile {$\mathcal{N}(h_0)$ or $\mathcal{N}(h_1)$ is not $B_I$-smooth}
	\EndFor
	\State We have initiated the descent tree with all leaves corresponding to $q < B_I$
	\State (Traditional special-$\mathfrak{q}$ descent)
	\State For each remaining prime factor $q$
		   in the factorization of $\mathcal{N}(h)$ we descend as follows:
	\State Set an initial upper bound $Q \gets q$
	\While {There exist primes $q$ in the descent tree of unknown log}
		\For {prime factors $\mathfrak{q}$ of $h$}
			\While {We have not found a $Q$-smooth relation}
				\State Construct the sieving lattice $L$ as shown in section 3.
				\State Reduce $L$ with LLL, giving $\overbar{L}$.
				\State Perform lattice sieving with $\overbar{L}$
				\While {There are unexamined relations remaining}
					\If {the norms on each side of the relation factor into primes
						   less than $Q$}
						\State exit while loop
					\EndIf
				\EndWhile
				\If {a $Q$-smooth relation was found}
					\State exit while loop
				\Else
					\State (very rarely) increase the allowable smoothness bound $Q$
				\EndIf
			\EndWhile
			\State An upper bound on primes to descend should now be less than $Q$
			\State Update $Q$ with the smaller bound
			\If {$Q < B$}
				\State Deduce the virtual log of $\mathfrak{q}$ from the log database
			\EndIf
		\EndFor
	\EndWhile
	\State All virtual logs of primes in the descent tree are now known
	\State We have $\text{vlog}(tg^i) = \text{vlog}(t) + i\text{vlog}(g)$
	\State Return $\text{vlog}(t)/\text{vlog}(g)\,\mod{\ell}$
\end{algorithmic}\label{alg4}
\algcomment{$\dagger$ Not strictly yet an algorithm as it is not proved to terminate.}
\end{algorithm}
\end{minipage}}

\section{Descent}\label{Descent}
Usually there are two parts to the descent step (also known as the individual
logarithm step) for relating the virtual logs of an
arbitrary generator/target to the known logs of the factor base.  These are
the initial split (or boot) step, followed by the special-q descent step.
The goal of the initial split step is to find an element in the finite field
whose lift to one number field has norm which decomposes into prime factors
which are `not too big', meaning that we are able to complete the descent by
using these primes as special-qs in the second step.  The special-q step
attempts to eliminate larger special-qs by expressing their virtual logs in terms
of virtual logs of smaller primes, and so on, until we reach the primes with
known logs in the factor base.\\

We might have to use a specialized algorithm for the initial split, depending on
the context.  For example, in large characteristic the continued fraction
algorithm is used.  New methods were described in \cite{GuillevicDescent} in the
extension field setting and we made use of these advances in our record
computation.\\

The algorithm from \cite{GuillevicDescent} for initial split makes use of the fact
that the target element will have the same log if it is multiplied by
any element in a proper subfield of $\F_{p^n}$ (thus it only applies
when $n > 1$).  Using this fact we can set up a lattice and the reduced
lattice gives us an element whose lift has smaller norm than that of the lift of
the target element, yet has the same virtual log.  The descent in extension
fields would be very difficult or practically impossible without a method such
as this to provide a starting point for the special-q descent.\\

\subsection{Special-q descent}

As already mentioned, in the special-q descent phase we recursively traverse the 
descent tree until we have reached elements of the number field whose norms
factor into prime ideals in our factor base only, for which we have a complete
(or near complete) set of virtual logs.  This process is basically the same
as the sieving step of NFS/exTNFS except in practice we must allow larger special-q primes.
For every prime ideal with an unknown log at a given leaf in the descent tree,
we use it as a special-$\mathfrak{q}$ and conduct traditional lattice sieving,
using the method shown in section \ref{section3}.
The idea is that at each subsequent leaf the primes get smaller.
We use a heuristic that the process must eventually terminate, but it is an open problem
to prove this rigorously.  The idea is not new, but since it is rare to see a formal
description of special-$\mathfrak{q}$ descent in the literature, we give pseudocode for
the process in `algorithm' \ref{alg4}.  Note however we cannot describe it as an algorithm
without a proof that it terminates.  In the case of index calculus using function fields
to solve DLP in e.g. small characteristic finite fields or in the Jacobian of a low degree
plane curve, there is some more detailed discussion of the descent
step, see for example \cite{indexcalculusdescent1}, which is referenced by
\cite{indexcalculusdescent2}.

\subsection{New method: Descent with intermediate random vectors}

In our record computation in $\F_{p^4}$ we implemented Guillevic's algorithm.
However, even with this algorithm we still found it difficult
to find a sufficiently smooth initial split.  We decided on a compromise - 
we allowed larger factors than usual in the initial split
algorithm, and fed these into an `intermediate' special-q step.\\

\noindent The idea is that we take one of the large initial split primes $q$ and
compute the sieving lattice $L$ as described in the previous section.
However, instead of sieving with this lattice (which would be impractical
since the associated norms would be too large to factor using the
primitive implementation of elliptic curve factoring in our 4d sieve program),
we compute random elements of the lattice, and if
the associated norm is `sufficiently smooth' (as determined by
PARI/GP's much more capable suite of factoring algorithms, with a set time
limit), we then complete the descent of
that $q$ with the usual special-q descent.  This simple idea made an otherwise
very difficult descent step much easier.  In fact the intermediate
randomised vector descent described can be implemented in a single line of PARI/GP 
and even without optimization works perfectly well.
Pseudocode for this algorithm is given in algorithm \ref{algdescentrandom}.

\begin{remark}
What distinguishes algorithm \ref{algdescentrandom} from lattice sieving is that there is no `sieve',
i.e. we just keep the special-q aspect and compute random vectors in the lattice,
which we show in the sequel has a good likelihood to eventually result in a
sufficiently smooth element of the number field.
\end{remark}

\subsection{Analysis of intermediate random vector step}

We first define the successive minima of a lattice $\Lambda$.

\begin{definition}
Let $\Lambda$ be a lattice of rank $n$. For $i \in \{1,\dots,n\}$ we define the $i$-th
successive minimum as
\[
\lambda_i(\Lambda) = \text{inf }\{r \vert \text{dim}(\text{span}(\Lambda \cup \overline{B}(0,r))) \geq i\}
\]
where $\overline{B}(0,r) = \{x \in \R^m \vert \lVert x \rVert \leq r \}$
is the closed ball of radius $r$ around 0.
\end{definition}

\noindent We state a famous result of Minkowski \cite{Cassels}, which we need below.

\begin{theorem}[Minkowski's Second Theorem]\label{Minkowski2}
For any full-rank lattice $\Lambda$ of rank $n$,
\[
\left(\prod_{i=1}^n \lambda_i(\Lambda)\right)^{1/n} \leq \sqrt{n}(\text{det }\Lambda)^{1/n},
\]
where $\lambda_i$ denotes the $i$-th minimum of $\Lambda$.
\end{theorem}

\subsubsection{Smoothness probability}

We would like to make the analysis of the likelihood of success of finding
a good initial split for a given field size more rigorous.  We reduce this
problem to an estimation of the likelihood that an integer of a given size
will be $y$-smooth, for some smaller integer bound $y$.  To this end we
have a theorem due to Canfield/Erd\H{o}s/Pomerance from the 1980s when
subexponential algorithms for factoring and discrete log were first being 
investigated in-depth.

\subsubsection{Canfield-Erd\H{o}s-Pomerance}\label{CEP}

\noindent We follow the presentation in \cite{CrandallPomerance}. 
A $y$-smooth integer has only primes at most $y$ in its prime factorization. Let
\[
\psi(x,y) = \#\{1 \leq n \leq x : n \text{ is }y\text{-smooth}\}.
\]

\noindent We have the following theorem:

\begin{theorem}
For each fixed real number $u > 0$, there is a real number $\rho(u) > 0$ such that
\[
\psi(x,x^{1/u}) \sim \rho(u)x.
\]
\end{theorem}

\noindent The function $\rho$ is known as the Dickman Rho function.  We then have the following
theorem due to Canfield, Erd\H{o}s, Pomerance:

\begin{theorem}\label{theoremCEP}
Let $\psi, \rho$ be defined as above.  Then
\[
\psi\left(x,x^{1/u}\right) = x u^{-u+o(u)}
\]
uniformly as $u \rightarrow \infty$ and $u < (1-\epsilon)\text{ln x}/\text{ln ln }x$,
for any fixed $\epsilon > 0$.
\end{theorem}

\noindent This gives a reasonable estimate for $\psi(x,y)$ when $y > \text{ln}^{1+\epsilon}x$
and $x$ is large.  Thus we can say the probability that an integer of size
$x$ is $x^{1/u}$-smooth is approximately
\[
x u^{-u+o(u)} / x = u^{-u+o(u)}.
\]

\noindent We can re-phrase this theorem using L-notation.  The
probability that an integer of size $L[a,\alpha]$ is $L[b,\beta]$-smooth is
\[
L\left[a-b,-\dfrac{\alpha}{\beta}(a-b)(1+o(1))\right].
\]

\begin{algorithm}[H]
\caption{Intermediate descent by random vectors (4d case)} 
\begin{algorithmic}[1]
	\Require polynomials $f, g$, tower polynomial $h$ defining exTNFS diagram
		\Statex large special-q prime q to descend
		\Statex $r,R$ such that $h(r) \equiv 0 \text{ mod } q, f(R) \equiv 0 \text{ mod } q$
		\Statex smoothness bound $B$
		\Statex coordinate interval size s (even integer for simplicity)
		\Statex factor time limit $t$ seconds
		\Statex overall time limit $T$ seconds
	\Ensure list of lattice vectors of increasing corresponding norm quality
    \vspace{1mm}
	\State $L \gets
		\begin{bmatrix}
		q & r & R & 0 \\ 0 & 1 & 0 & R \\ 0 & 0 & 1 & 0 \\ 0 & 0 & 0 & 1
		\end{bmatrix}$
	\State reduce $L$ with the LLL algorithm yielding lattice basis $L'$
	\State $Q_{best} \gets q$
	\While {true}
		\State $v \gets (rand(s) - s/2,\dots,rand(s) - s/2)$ with 4 entries
		\State $(a,b,c,d) \gets L' \cdot v$
		\State $N_1 \gets \text{Res}_y(\text{Res}_x(a+by+(c+dy)x,f),h)$
		\State $N_2 \gets \text{Res}_y(\text{Res}_x(a+by+(c+dy)x,g),h)$
		\State $Q_1 \gets \text{max}\{q, \text{largest prime in factorization of }N_1\text{ in time }t\}$
		\State $Q_2 \gets \text{max}\{q, \text{largest prime in factorization of }N_2\text{ in time }t\}$
		\If {$Q_1 < Q_{best} \text{ }\mathbf{ and }\text{ } Q_2 < Q_{best}$}
			\State $Q_{best} \gets \text{max}\{Q_1,Q_2\}$
			\State record $(a,b,c,d)$, also output to screen
		\EndIf
		\If {time limit $T$ exceeded}
			\State exit while and return best $(a,b,c,d)$
		\EndIf		
	\EndWhile
\end{algorithmic}\label{algdescentrandom}
\end{algorithm}

\subsubsection{Complexity of algorithm \ref{algdescentrandom}}

\noindent We present the following theorem to analyse the success likelihood of descent
using random vectors.

\begin{theorem}\label{randvecdescent}
Given a target smoothness bound $B \in \Z$ and a coordinate interval of size $s$,
the asymptotic probability of success
of algorithm \ref{algdescentrandom} in dimension 4 is $u^{-u}$, where
$u = \text{log}(pqs^4)/\text{log}(B)$.
\begin{proof}\normalfont
By Minkowski's Second Theorem, since the determinant is $q$, short elements of the reduced
4d lattice in algorithm \ref{algdescentrandom} have coordinates bounded by $O(q^{1/4})$.
The matrix vector product then gives vectors with coordinates of size $O(sq^{1/4})$.
Suppose the finite field element $A = a+by+(c+dy)x$ lifts to an element $\alpha$ in
one of the tower number fields $L$. We take the double resultant with the tower field
polynomials $h$ and $f$ from the conjugation construction where
\begin{alignat*}{4}
h &= y^2 + h_1y + 1,      &h_1 \in \Z, h_1 &= O(1)\\
f &= f_0x^2 + f_1x + f_0, &f_i \in \Z, f_i &= O(p^{1/2}).
\end{alignat*}
The double resultant $\text{Res}_y(\text{Res}_x(A,f),h)$ has coefficients of terms in
$\{f_0,f_1\}$ of degree 2 multiplied by terms in $\{a,b,c,d\}$ of degree 4.
Thus, since $f_i = O(p^{1/2})$ and $a,b,c,d = O(sq^{1/4})$,
$\alpha$ will have norm $O(pqs^4)$.  Applying the theorem of Canfield/Erd\H{o}s/Pomerance
(see section \ref{CEP} above), we expect to have to trial approximately $u^u$ integers of size
$x$ to find one which is $x^{1/u}$-smooth.  Since our norms are of size $O(pqs^4)$, we
need $B = (pqs^4)^{1/u}$, so we should set $u = 1/\text{log}_{pqs^4}(B) = \text{log}(pqs^4)/\text{log}(B)$.
\end{proof}\qed
\end{theorem}

\begin{remark}
Concretely, for our $\F_{p^4}$ computation, $p$ has 128 bits,
and we could find from the initial split an element with norm prime factors up
to say 70 bits.  Taking such a maximum prime, we use this as the special-q.
The coordinate interval size $s$ can be quite small, as long as our pool of random
vectors is large enough.  We can take $s = 200$.
Thus $pqs^4$ will have 229 bits.  Say we want another element which has
up to at most 50-bit primes, so that our lattice sieve can complete the descent
more easily.
This means we should take $u = \text{log}(pqs^4)/\text{log}(2^{50}) \approx 4.58$, and so we
expect to have to try $u^u \approx 1064$ norms until we find such a split.
Of course this could be slightly higher (or slightly lower), because of the $o(u)$ in the
theorem of Canfield/Erd\H{o}s/Pomerance, but not by too much - we are
in an acceptable range of order of magnitude.
\end{remark}

\noindent We offer another perspective using L-notation by estimating a lower bound
for the running time of algorithm \ref{algdescentrandom} using a lemma from 
\cite{GuillevicDescent} which we paraphrase here:

\begin{lemma}[Running-time of $B$-smooth decomposition]
Let $Q = p^n$ (in our concrete example $n = 4$).
Assume that the norm $N$ of the random elements in $\F_Q$ in
algorithm \ref{algdescentrandom} is bounded by $Q^e = L_Q[1,e]$.  Write
$B = L_Q[\alpha_B, \gamma]$ the smoothness bound for $N$.  Then the lower bound
of the expected running time for finding $B$-smooth $N$  is $L_Q[1/3,(3e)^{1/3}]$,
obtained with $\alpha_B = 2/3$ and $\gamma = (e^2/3)^{1/3}$.
\end{lemma}

\noindent Using this lemma we only have to compute $e$ and we are done.
The size of our norms $N$ is $pqs^4$ using the notation of theorem \ref{randvecdescent}.
We require that $Q^e = pqs^4$, so we should take $e = \dfrac{1}{4}\log{pqs^4}/\log{p}$.

\begin{remark}
We could spend more time considering the $o(u)$ in theorem \ref{theoremCEP} and its
effect on the running time of algorithm \ref{algdescentrandom} but in practice the 
algorithm is extremely fast, as we discuss below.  We have at least
a theoretical asymptotic lower bound on the runtime using the above discussion.
\end{remark}

\subsection{Performance of intermediate descent}

Algorithm \ref{algdescentrandom} is extremely fast in practice, requires
minimal memory (since there is no sieve) and can be arbitrarily parallelised.
The speed is only limited by the time to factor the norms that arise from
the given special-q.  We can set a time limit on the factorization which
allows spending more time factoring promising candidates and ignoring
hard-to-factor norms (these will take longer to factor and this can effectively
be detected by setting a threshold factorization time).  Running the algorithm on one core
for only a couple of minutes was sufficient to descend a large special-q prime of 68 bits
down to primes of at most 52 bits.  Without this intermediate step, our implementation
of Guillevic's 4d descent algorithm from \cite{GuillevicDescent} set to find a 64-bit smooth norm
did not find any suitable split even after an appreciable core hour cost.

\section{Record Computation in $\F_{p^4}$}\label{Record}
We give details of our record computation in $\F_{p^4}$ for a field size of 512 bits.
The previous record in $\F_{p^4}$ was 392 bits, using the original number field sieve
\cite{PrevRecFp4}.  All core timings here are normalised to a nominal 2.1GHz clock speed.

\subsection{Polynomial Selection}
We selected the prime
$p = 314159265358979323846264338327950288459 = \lfloor 10^{38}\pi\rfloor + 40$.
This was so that $\Phi_4(p) = p^2+1$ has a large prime factor
$\ell = (p^2+1)/28047119146$.  Our target field is $F = \F_{p^4}$ and we
want to solve discrete logs in the subgroup of $F$ of size $\ell$.
We used the conjugation method of polynomial selection described in section 2
to produce the polynomials
\begin{align*}
h &= y^2 - y + 1 \in \Q[y]\\
f_0 &= 2690013449567156494\alpha x^2 - 3386516025263921869x + 2690013449567156494\alpha \in K[x]\\
g_0 &= (\alpha - 1)x^4 + (2\alpha - 47)x^2 + (\alpha - 1) \in K[x]
\end{align*}
with
\[
\text{gcd}(\overbar{f_0},\overbar{g_0}) = \overbar{f_0}
\]
and we have the number fields
\begin{align*}
K &= \Q(\alpha) \cong \Q[y]/\langle h \rangle \\
L_0 &= K(\beta) \cong K[x]/\langle f_0 \rangle \\
L_1 &= K(\gamma) \cong K[x]/\langle g_0 \rangle.
\end{align*}

\subsection{Sieving}
We implemented our 4d sieve in C\texttt{++} using g\texttt{++}, the Gnu C\texttt{++} compiler.
We kept the approach of list/sort from \cite{MR4178778} for identifying
cofactorization candidates.  The parameters for sieving were as
follows:  We set a large prime bound of $2^{26}$.  This means the
factor base contains $24,735,135$ ideals of degrees $1,2,4$ on either side.
We are mainly interested in degree 1 ideals only and in total there are
$7,913,047$ of these in the factor base, but we allow a small number of
degree 2 ideals whenever they show up.  Our sieve base bound (i.e. a bound
on the primes used for sieving only) was $4,000,000$.  We set a
bounding orthotope for 4d sieving of dimensions $64\times64\times64\times64$.
Our special-qs were in intervals between $4,000,000$ and $67,108,864$.
We spent $52,663$ core hours collecting $58,793,420$ raw relations
(but including exact duplicates the 4d sieve found $ 75,192,614$ relations).
This was done over 48 hours on 24 nodes of the Kay cluster at the
Irish Centre for High-End Computing (ICHEC).  Each compute node
has 40 cores.

\subsection{Duplicate Removal}
Duplicate removal is necessary to avoid problems in the linear algebra step.
As described in \cite{GDeMthesis}, in the TNFS setting there are
a number of different types of duplicate relation.  The main
criterion we used was to enforce keeping only one relation
$\langle a+b\alpha+(c+d\alpha)\beta\rangle$ whenever $(a+by)/(c+dy) \mod{\overbar{h}}$ is the same
between two or more relations.  We found that also requiring
$\text{gcd}(N_0,N_1) = 1$ for norms $N_0, N_1$ on sides 0,1 respectively
threw away too many relations so we allowed any gcd.  Ensuring
a surplus of relations vs ideals seems to be a very delicate matter.
If we sieve too little and carry out duplicate removal, it can happen
that too many relations are removed and we end up with less relations
than unknowns.  Even erring on the side of caution, when we removed
duplicates from the set of $58,793,420$ raw relations, we were left
with $6,619,495$ unique relations involving $6,606,889$ distinct
ideals - barely a surplus but it was enough to proceed.
The duplicate removal step took 1,180 core hours, however our
program for doing this could be significantly optimized.

\subsection{Matrix Preprocessing}
We used the CADO-NFS \cite{CADO} binaries purge and merge-dl
from the development branch (commit 0x03048880f12f8...) to reduce
the size of the matrix passed to the linear algebra step.  The
purge binary removed singletons and reduced the relation set to
$3,552,319$ relations.  Then merge-dl generated a matrix of size
$1,093,016$ rows and $1,093,012$ columns, allowing for 1 Schirokauer
map on side 0 and 3 Schirokauer maps on side 1.

\subsection{Schirokauer Maps}
We wrote our own program to generate the Schirokauer map columns.
We note that we used an isomorphism between the relative number
field on a given side, and its absolute number field.  This isomorphism
was computed with Sage \cite{sagemath}.  Then Schirokauer maps are generated in
the expected way.  Our program uses libpari, the pari-gp library
\cite{PARI2} for number field/multivariate polynomial arithmetic.

\subsection{Linear Algebra}
We used CADO-NFS's implementation of the Block Wiedemann algorithm
to find a right nullspace vector for our linear system.  The matrix
was small enough to be able to process in just over 3 days of real
time on a 16 core home workstation.  This step took $1,187$ core hours.

\subsection{Log Reconstruction}
Since we chose the factor base format, we also wrote our own log
reconstruction program.  It works in a similar way to the CADO-NFS
reconstructlog-dl binary, by multiple passes where on input of
the total set of relations, including those removed by purge,
the program deduces logs from relations with a single unknown.
This allows further passes where now more relations with a single
unknown are present, and continues until there are no more deductions
possible.  With our program we reconstructed $6,160,438$ known logs
in 29 passes.

\subsection{Descent}
Working in the subgroup of size
\[
\ell = 3518936953814357579166997631392367151668364387422300934981051190217
\]
of $F_{p^4}$ defined as above, we chose the generator
\[
g = x + 5
\]
and target, from the digits of exp(1),
\begin{equation*}
\begin{split}
t = (27182818284590452353602874713526624977y + \\
57247093699959574966967627724076630353)x + \\
(54759457138217852516642742746639193200y + \\
30599218174135966290435729003342952605).
\end{split}
\end{equation*}
The initial split step for the target produced an element of the number field $L_0$ of norm
\begin{equation*}
\begin{split}
&2^2 \cdot 5701 \cdot 41611 \cdot 55057 \cdot 4088911 \cdot 996853403317 \cdot 203630288936359\phantom{.}\cdot \\
&512871673683067 \cdot 1796070586527211 \cdot 247959619100557519 \cdot 244801552463017277719
\end{split}
\end{equation*}
where the largest prime has 68 bits.  The initial split step/intermediate descent
took about 12 core hours, and computing
the special-q descent trees with sieving took about 12 hours of manual effort.
We had two versions of our 4d sieve program - a 32-bit version, used in the
sieving step, and a 64-bit version which was slower, but necessary for larger
primes in the descent tree.\\

Finally, we computed the virtual log
\[
\text{vlog(g)} = 992323251125728356329649930303177107284104491653542204374572554143
\]
of the generator and
\[
\text{vlog(t)} = 401809551984744589507112134228751535116674975282792047359473327871
\]
of the target and it can be confirmed easily that
\[
g^{C \cdot \text{vlog}(t)} \equiv t^{C \cdot \text{vlog}(g)}.
\]
where $C = (p^4-1)/\ell$.

\section{Conclusion}
We implemented the key components of the Extended Tower Number Field Sieve and together
with linear algebra components of CADO-NFS demonstrated a total discrete log break
in a finite field $\F_{p^4}$ of size 512 bits, a new record.  This provides another
data point in the evaluation of security of systems dependent on the intractability
of discrete log attacks in extension fields.  Whereas the recent articles
\cite{GDeMthesis}, \cite{GDeMpaper} show that asymptotically, sieving in a d-dimensional
sphere is optimal as $d \rightarrow \infty$, there seems to be room at lower
dimensions for sieving in an orthotope to remain competitive.  We did not optimize our code
particularly well and there are probably further gains in sieving speed possible.
The parameters for sieving were
tuned in an ad-hoc way and a finer examination of optimal parameters would be interesting.
It would be a fairly easy change to adjust the sieving shape to best suit a given
special-$\mathfrak{q}$, for rectangular sieving orthotopes, improving the relation yield.
Finally, we did not exploit the common Galois automorphism of the sieving polynomials,
which would have cut the sieving time in half.
The overall timings of the key stages of our computation are shown in
table \ref{timings}.

\begin{table}[!ht]
\begin{center}
\begin{tabulary}{17cm}{|C|C|}
\hline
\T Stage&Time Taken (Core Hours)\\
\hline
\T Relation Collection&52,663\\
\hline
\T Duplicate Removal&1,180\\
\hline
\T Linear Algebra&1,187\\
\hline
\T Descent&24\\
\hline
\T Total&55,054\\
\hline
\end{tabulary}
\end{center}
\caption{Overall Timings}\label{timings}
\end{table}

\vspace{-5mm}

\nocite{*}
\bibliographystyle{splncs04}
\bibliography{refs}

\end{document}